\documentstyle[twoside,fleqn,epsf,espcrc2]{article}

\newcommand{\AmS}{{\protect\the\textfont2
  A\kern-.1667em\lower.5ex\hbox{M}\kern-.125emS}}

\newcommand{\gsim}{{\protect
  \kern.18em\lower.5ex\hbox{$\stackrel{>}{\sim}$}\kern.25em}}
\newcommand{\lsim}{{\protect
  \kern.17em\lower.5ex\hbox{$\stackrel{<}{\sim}$}\kern.23em}}

\newcommand\be{\begin{equation}}
\newcommand\ee{\end{equation}}
\newcommand\bea{\begin{eqnarray}}
\newcommand\eea{\end{eqnarray}}
\newcommand{\psb}{\bar{\psi}}

\title{
{\vspace{-3cm} \normalsize
\hfill \parbox{30mm}{DESY 95-141}}\\[25mm]
Recent Developments in Fermion Simulation Algorithms         
        \thanks{
                Plenary talk given at the International
                Symposium on Lattice Field Theory, 4-8 June 1996,
                St. Louis, Mo, USA}}

\author{K. Jansen \address{Deutsches Elektronen Synchroton, DESY,
        Notkestr. 85, \\ 
        22603 Hamburg, Germany}%
        \thanks{e-mail: kjansen@desy.de}}
           
\begin{document}

\begin{abstract}
A summary of recent developments in the field of 
simulation algorithms for dynamical fermions is given.
\end{abstract}

\maketitle

\section{Introduction}

Since Kennedy's complaint \cite{tony} that there has been little
activity and progress in the field of fermion simulation
algorithms, we have seen much work devoted to this 
algorithmic challenge lattice field theory has to face.  
The most prominent one is  
L\"uscher's suggestion \cite{martin} to use a reformulation of
the QCD partition function in terms of a number of       
bosonic field copies.
This multiboson technique to simulate dynamical fermions 
has been studied thoroughly in the last years
\cite{bunk,beat,peardon,forcrand1,forcrand2,forcrand3,forcrand4,kramersboson,borici} 
and it was established to lead to
an {\em exact} algorithm which is competitive to the so far
most often used Hybrid Monte Carlo (HMC) algorithm \cite{hmc} or
its variant, the Kramers equation or L2MC algorithm 
\cite{horowitz,julius,italiens,kramers}.   

On the other hand, with the introduction of the multiboson technique,
a renewed interest to accelerate also the HMC algorithm was stimulated.
Better preconditioner, as discussed by Frommer at this conference
\cite{andreas}, improved integration schemes and alternative choices of
matrix inverters led to a considerable amount of progress. On the negative
side possible problems
with the reversibility condition of the HMC algorithm has been 
encountered 
\cite{revers,kramers,wuppertal1,brower,horvath,liu}.  
Finally, simulations of 
{\em dynamical} fermions with Symanzik improved
actions have already been started.  

In this review talk 
fermion simulation algorithms for lattice QCD in the Wilson
formulation are studied. 
For lack of space, I have to refer to e.g. \cite{kramersboson} for the
necessary notations and definitions.  

\section{The algorithm race}

The question one is
presumably most interested in, is,  
which of the partly conceptually very different but exact 
algorithms is the fastest in the sense that an
independent configuration is generated within a given amount of
computer time. To find an answer, we will compare particular
implementations of the multiboson technique against the HMC or
the Kramers equation algorithm. 

The last mentioned algorithm is a variant of the HMC algorithm. In the
continuum the fictitious time evolution is described by a set of formal 
stochastic differential equations in the fields $\phi$ and their 
conjugate momenta $\pi$, 
\be \label{kramers}   
\dot{\pi}  = -\frac{\partial H}{\partial \phi} -\gamma
\pi+ \eta(t)\; ;\;  \dot{\phi} = \pi\; .
\ee  
In eq.(\ref{kramers}) $H$ denotes the Hamiltonian and 
a friction term with a coefficient 
$\gamma$ appears that can be tuned to optimize the algorithm. 
A Gaussian noise $\eta$ is fed into the time evolution of the system
at each time step. 
A recent free field analysis of generalized
molecular dynamics algorithms like the Kramers equation 
has been done in \cite{freekennedy}.  
In the case of Wilson fermions the performance 
of the Kramers equation is comparable   
to the HMC algorithm \cite{kramers}. For staggered fermions the 
situation seems to be more favorable for the HMC algorithm \cite{meyer},
although no definite conclusion can be given at the moment.

\subsection{Scaling of the algorithms}

In order to compare the performance of the algorithms one would like
to understand their scaling behaviour better. In this review, I will
choose the lowest eigenvalue $\lambda_{min}$ of $Q^2$ (
$\sim M^\dagger M$ with $M$ the Wilson-Dirac operator) 
and the volume
$V$ of the lattice as the relevant scaling variables. 
Note that for staggered fermions the corresponding $\lambda_{min}$
would be directly proportional to the quark mass squared. 
The choice of $\lambda_{min}$ 
seems to be natural if one has the Schr\"odinger functional (SF) 
formulation \cite{sf} in mind, where 
simulations at zero quark masses
are possible while $\lambda_{min}$ remains non-zero \cite{letter,paper}.
        
\vspace{0.1cm}
{\bf \noindent number of fields/CG iterations}\\  
\noindent The accuracy $\delta$  
of the (Chebyshev) polynomial $P$ approximation
in the multiboson technique is exponential, 
$ \delta \sim \exp\{-2\sqrt{\epsilon}n\}$ \cite{martin}. 
Since $\epsilon$ has to be chosen according to the value of $\lambda_{min}$,
in order to have a fixed accuracy the number of boson fields 
$n \sim 1/\sqrt{\epsilon} \sim 1/\sqrt{\lambda_{min}}$.

The number of fields as a function of $\lambda_{min}$
can be compared to the number of iterations
needed in the molecular dynamics algorithms to invert the matrix
$Q^2$. Fig.~1 shows 
the {\em time} per call of the matrix inversion routine 
in --for this discussion-- irrelevant units 
as a function of the condition number
$k=\lambda_{max}/\lambda_{min}$, with $\lambda_{max}$ the largest
eigenvalue of $Q^2$. 

\vspace{-5mm}
\begin{figure}[htb] \label{figure1}  
\centerline{ \epsfysize=6.5cm 
             \epsfxsize=6.5cm 
             \epsfbox{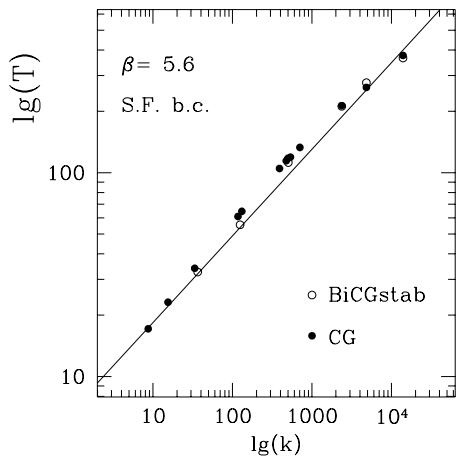}}
\vspace{-5mm}
\caption{Time spent in the matrix inversion routine for the CG (full circles)
         and the BiCGstab (open circles).}
\vspace{-9mm}
\end{figure}

For the inversion the CG and the BiCGstab \cite{wuppertal2} 
methods have been
used. The double logarithmic plot shows that both 
methods are compatible with a linear
behaviour and that they basically
require the same amount of computer time for the matrix inversion. Note 
that the condition number is ranging over several orders of magnitude.
The slope of the straight line in fig.~1
is roughly $0.43$ and
therefore somewhat better than the expected behaviour \cite{scales}.  

An important remark is that for a smaller value of                     
$\beta =5.4$ we find, in accordance with \cite{wuppertal2},
that the BiCGstab shows a gain of about 15-20\% in spite of the fact
that Schr\"odinger functional boundary conditions 
have been used.
The observation that at $\beta=5.6$ a comparable performance is seen,
indicates that the competetivity of the two methods depends on the
parameters chosen.

\vspace{0.1cm}
{\bf \noindent autocorrelation times}\\
\noindent Little to nothing is known about the scaling of the 
autocorrelation time with
respect to $\lambda_{min}$. I take therefore the freedom
of having an optimistic point of view and adopt the free field
results for all the algorithms. This amounts to have an autocorrelation
time $\tau \sim 1/\sqrt{\lambda_{min}}$. For the multiboson technique
one expects this behaviour for an optimized Hybrid-over-relaxation
algorithm. For the HMC algorithm the same behaviour is found 
in the limit of long trajectories \cite{kenpen}. 
The Kramers equation algorithm assumes this with an optimal
choice of $\gamma$ for a single step   
trajectory \cite{horowitz}.

\vspace{0.1cm}  
{\bf \noindent additional scaling factors}\\
\noindent Both type of algorithms show, unfortunately, an additional
dependence on the lowest eigenvalue. For the multiboson technique
it is by now well established that the autocorrelation time has a
linear dependence on the number of boson fields and therefore 
$\tau \sim n \sim 1/\sqrt{\lambda_{min}}$ 
\cite{beat,peardon,forcrand1,forcrand4}.  

In the molecular dynamics kind of algorithms it is the tuning of the
step size $\delta\tau$ to keep a constant acceptance that gives
rise to an additional scaling factor. One finds,   
assuming an integrator with $\delta\tau^3$ error \cite{gupta}
that $\delta\tau \sim 1/\lambda^{3/4}_{min}$ for the HMC algorithm and
by similar arguments $\delta\tau \sim 1/\sqrt{\lambda_{min}}$ for the
Kramers equation algorithm. Using improved integration schemes will,
of course, soften this behaviour. 

\vspace{0.1cm}
{\bf \noindent volume dependence}\\
\noindent For the multiboson technique one expects a volume $V$ behaviour
as $V(logV)^2$ \cite{forcrand1,forcrand4}. The $(logV)^2$ appears
in the exact version of the algorithm that is discussed below.
To have a constant acceptance in this algorithm, the accuracy
$\delta \sim 1/\sqrt{V}$. Therefore the number of fields $n \sim logV$
and since the autocorrelation time scales with $n$ we finally find the
$(logV)^2$ behaviour.

The volume dependence in the molecular dynamics kind of algorithms
comes basically from the acceptance behaviour. The free field analysis
\cite{kenpen} gives  $P_{acc} \sim erfc(N_{md}\delta\tau^3\sqrt{V})$, 
where $N_{md}$ is the number of molecular dynamics steps.
From this it follows that the HMC algorithm with $\delta\tau N_{md} =1$
gives a $VV^{1/4}$ behaviour whereas the Kramers equation algorithm
with $N_{md}=1$ should give $VV^{1/6}$. 
%
For a fixed acceptance rate one finds in practise 
that the scaling of the step size 
does not contradict the theoretical estimate.

\subsection{Tune up of the algorithms} 

There are several improvements that can be done to accelerate the
``bare'' algorithms. The most important of these is preconditioning
the fermion matrix. 

\vspace{0.1cm}  
{\bf \noindent preconditioning}\\
\noindent Using the standard even/odd preconditioning 
\cite{degrand} one not only finds that the lowest eigenvalue is 
increased by a factor of roughly four but that at the same time
the largest eigenvalue is lowered by a factor of about two.
This leads to a factor of about 8 improvement for the condition number.

The consequence for the multiboson technique is that the number of
fields is decreased whereas in the molecular dynamics algorithms
the number of iterations for the matrix inversion is substantially
reduced. 

\vspace{0.1cm}  
{\bf \noindent update}\\
For the update parts of both simulation methods it turns out that
for the multiboson technique one has to optimize the mixing ratio
of heatbath and over-relaxation sweeps \cite{beat,forcrand4}. In
the case of the molecular dynamics algorithm a better integration scheme
as suggested by Sexton and Weingarten \cite{sexy} gives considerable
improvement \cite{kramers}. 

\vspace{0.1cm}  
{\bf \noindent matrix inversion}\\
At the time of the conference a comparison between an exact version
of the multiboson technique in its hermitian and 
non-hermitian variant is missing. Therefore
no definite conclusion about which polynomial to choose can be given.

On the side of the molecular dynamics algorithms, it seems that the
CG and the BiCGstab methods show a comparable performance for the
inversion of $Q^2$ whereas
the minimal residual is doing somewhat worse. Using a higher order
integration scheme allows to choose larger step sizes. It seems therefore
to me that for present applications there is no real need to use the
chronological extrapolation method \cite{brower}. If, on the other
hand, one reaches situations where the step size becomes small, this
way of finding a good starting vector might become relevant.

It seems that for the inversion of $Q^2$ the CG algorithm is 
optimal. Therefore the only improvement on this side may come from
better preconditioning techniques like the Oyanaga ILU preconditioning
in combination with the Eisenstat trick \cite{ssor}.  
An interesting idea is to use this to precondition the 
{\em fermionic force} \cite{takaishi} which would allow for larger step 
sizes.
It was suggested that one might increase the stopping criterion of the
inversion routine by several orders of magnitude \cite{takaishi}
from $\|r\|=10^{-8}$, a standard choice, to $\|r\|=10^{-3}$ which
still provides a reasonable acceptance rate. 
Of course, by doing this, one has to choose the starting vector for
the inversion to be always the same in order to guarantee the reversibility
of the algorithm. It is, however, 
unclear to me, whether the drastic change of
the residuum by five orders of magnitude will not lead to 
reversibility problems in practical simulations.

\subsection{Making the multiboson technique exact}  

There have been several proposals to make the multiboson technique
exact \cite{bunk,beat,peardon,forcrand1}. However, I consider the
suggestion in \cite{forcrand4} the most promising one.          
One can write the {\em exact} partition function with the local
bosonic action $S_b$ in the form  
\be   
Z = \int{\cal D} U\int{\cal D}\Phi^\dagger{\cal D}\Phi
|det (MP(M))|^2 e^{-S_g -S_b}\; .
\ee
The correction factor $|det (MP(M))|^2$ can be written in terms of new
Gaussian fields $\eta$,  
\be     
\int{\cal D}\eta^\dagger {\cal D}\eta
e^{-|[MP(M)|^{-1}\eta |^2} \equiv \int{\cal D}\eta^\dagger {\cal D}\eta
e^{-S_C}
\ee
which defines the correction action $S_C$. 
The next step is to impose an accept/reject step on the correction
action $S_C$. In order not to calculate the determinant factor wich
appears because we are considering probability densities, one has to
``order'' the gauge fields. We say that $U$ succeeds $U'$, 
$U \succ U'\;\; \mbox{if} \;\;\; S_g(U) \geq S_g(U')$.   

The probability $P=P_{\eta\rightarrow\eta'} (U,U')$ is then 
\be    
P=\left\{ \begin{array}{cc}
\frac{c}{|det (MP(M))|^2}e^{-S_C(U,\eta')}
& \mbox{if}\;
                                            U \succeq U' \\
\frac{c}{|det (M'P(M'))|^2}e^{-S_C(U',\eta')}
& \mbox{if}\;
                                            U \prec U' \; .
\end{array} \right.
\ee

This probability density guarantees that in the accept/reject step the
determinants cancel. In \cite{forcrand4} the detailed balance proof
of the above scheme is given.  
A nice feature of this approach is that one can estimate the 
acceptance rate analytically 
and 
the data seem to obey this analytical form in a convincing
manner \cite{forcrand4}.   

\subsection{Performance comparison}

The two kinds of algorithms have been compared for two situations. 
The first one
is for QCD with gauge group SU(2) \cite{kramersboson}. 
Here a hermitian polynomial was
chosen and no Metropolis test was employed. However, the algorithm
parameters were chosen such that the observables agree.
On the molecular dynamics side, the Kramers equation algorithm 
was used. The tests are performed at a rather large pion to 
$\rho$-mass ratio of $m_\pi/m_\rho =0.95$. On the other hand,
several lattice sizes $6^312, 8^312$ and $16^4$ were taken.
The second test has been performed by A. Galli, C. Liu and myself 
with SU(3) as the gauge group.
Here the multiboson technique had the exactness step implemented
and its non-hermitian version was chosen. The comparison was done
on a $8^4$ lattice at $\beta=5.6$ and $\kappa =0.1585$ which corresponds
to the critical value \cite{gupta2}.  
It is gratifying to see in this case 
that without any extrapolation the numbers of
the multiboson simulation come out to be completely consistent
with the HMC value.    

\begin{figure}[htb] \label{figure3}  
\centerline{ \epsfysize=6.5cm 
             \epsfxsize=6.5cm 
             \epsfbox{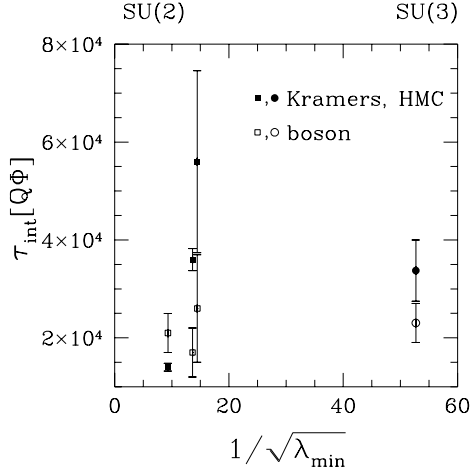}}
\vspace{-5mm}
\caption{Performance comparison for the multiboson technique and the
         Kramers equation (SU(2)) and the HMC (SU(3)) algorithms.}  
\vspace{-9mm}
\end{figure}

In fig.~2 the integrated 
autocorrelation times for the plaquette in more machine
independent units of matrix times vector,
$Q\Phi$, operations is shown for the two test cases. Given the
large error bars, one can say that the 
algorithms perform comparably. It is surprising
that the decrease of $\lambda_{min}$, when going from SU(2) to SU(3),
does not seem to increase the autocorrelation time.

\section{More participants in the race}

In this section I will shortly describe alternatives to the algorithms
described above.

{\bf \noindent bermions}\\
Bermion simulations \cite{bermions} are done at negative
flavour numbers $n_f$. 
This at first sight strange idea appears to be very
fruitful in practise. The major step forward in this approach is to
impose non-perturbative matching criteria to relate the different
negative quark flavours to zero and finally to, say, $n_f=2$.
During the course of investigating the bermion method,
the authors also found a nice way to compute propagators in terms of
pseudofermions.

The bermion approach can certainly serve to explore the parameter
space as they are much cheaper than a full QCD simulation. As such they
can give a very good first guess and guide the 
choice of parameters. Of course, at the end one would like
to perform a real simulation with positive $n_f$ in order to verify
the result. 
An open question in the bermion approach is, whether
there exists a value of the quark mass at which the bermion method
breaks down.  

{\bf \noindent domainwall fermions}\\
It was first pointed out in \cite{shamir} that the idea of
domainwall fermions, originally invented to shed light on the problem
of chiral fermions on the lattice, could also be used for simulations
of lattice QCD. At least for an infinite extent of the extra dimension,
required in the domainwall fermion approach, the quark mass gets only
a multiplicative renormalization. Tests in the vector Schwinger model 
\cite{jaster,vranas} seem
to indicate that in practise one might end up with only a moderate
number of slices of the extra dimension to approach the chiral limit.
It is, however, too early to say, whether domainwall fermions can
lead to a competitive method for full QCD simulations. 

{\bf \noindent Supersymmetry}\\
It is interesting to see that  supersymmetry 
with Majorana fermions in the adjoint representation 
is now attacked on the
lattice. The pioneering work by Montvay \cite{montvay} used the
multiboson technique to setup a simulation program. There, a particular
construction of the polynomial in a two step procedure to approximate
the Pfaffian was given. From the point of view of this review talk
it is intriguing that the results by Montvay have been recalculated 
\cite{hmd}   
by using a Hybrid molecular dynamics algorithm and a comparable
performance of both algorithms have been found. Of course, the Hybrid
molecular dynamics approach has errors $(\delta\tau)^2$ and one
would therefore prefer an exact version of the multiboson technique.

There are additional attempts like Slavnov's way of bosonization
of the fermion action \cite{slavnov} or the approach by Liu and Thron
to get rid of the pseudofermions altogether
\cite{thron}. But these works are at a 
very early stage and one has to wait in order to see their
benefits. The attempts to use adaptive step size methods seem not to lead to 
further progress \cite{taka}. A final but important remark is that 
with the multiboson technique simulations
with $n_f=1$ might be feasible \cite{nf=1}.  

\section{Reversibility}

In order for the HMC algorithm to be exact, the equations of motion
used therein have to be reversible. Although this is certainly the
case for a computer with infinite precision arithmetic, in daily
life one has to face rounding error effects. Indeed, it was
noticed that in practical simulations the reversibility condition
is not satisfied \cite{revers}. Moreover, in \cite{kramers} it was
pointed out that the equations of motion as used
in the HMC algorithm are chaotic in nature and have a positive
Liapunov exponent $\nu$. 

This implies that rounding errors eventually become exponentially amplified.
The values of the Liapunov exponents have been studied as a function
of the parameters $\beta$ and $\kappa$ \cite{liu,horvath}. 
Whereas the $\beta$ dependence is noticeable, the $\kappa$ dependence
appears to be rather weak. In \cite{horvath} it was suggested that
the Liapunov exponent scales like the correlation length $\xi$ such that
$\nu\xi = const$. One has to see, whether this interesting hypothesis
will withstand future tests. 

At the moment no quantitative estimate of the exponential
amplification of rounding errors can be given as far as physical
observables are concerned. However, estimates \cite{liu,lippert}
indicate that one might face problems on lattice sizes larger than
$32^4$ with 32 bit arithmetic. There the reversibility error 
might reach a level of several percent.

\section{Symanzik improvement}

During the conference, the use of improved actions has certainly been
a major topic. The investigations
of the effects of improvement have so far been restricted to the quenched
approximation. It is a natural and necessary next step, to implement
improved actions also for dynamical fermions. 

Here I want to concentrate only on the Symanzik 
on-shell improvement program
\cite{symanzik}. Following this program leads to the introduction
of a Sheikholeslami-Wohlert (SW) term as suggested in \cite{clover}: 
\be \label{cloveraction}  
S_{sw} =
      {i \over 4} c_{sw} \sum_{x,\mu,\nu}
      \psb(x) \sigma_{\mu\nu} {\cal F}_{\mu\nu}(x) \psi(x)\; .
\ee

For the implementation of the SW-term in the molecular dynamics
algorithms, HMC and Kramers equation, one needs to find the equations
of motion while preserving even/odd preconditioning. These equations
can straightforwardly be derived \cite{luo,sw_hmc,wingate}. 
A complete implementation of the SW-term in a 
dynamical fermion simulation program has been done in
\cite{sw_hmc}. There also possible tests of the code, using strong coupling
expansions for the plaquette and ${\cal F}^2$ are given. 
An independent complete implementation for the case of finite
temperature simulations has been done at SCRI \cite{wingate}.



\vspace{-10mm}
\begin{figure}[htb] \label{figure4}
\centerline{ \epsfysize=6.5cm
             \epsfxsize=6.5cm
             \epsfbox{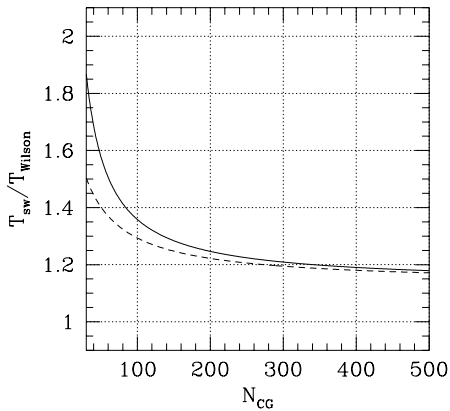}}
\vspace{-11mm}
\caption{Ratio of the time needed for the SW-action as compared 
         to standard Wilson fermions.}  
\vspace{-9mm}
\end{figure}

Of course, the crucial question is, whether the overhead of the
SW-term as compared to standard Wilson fermions appears to be
small enough to justify its use in practical simulations. 
In fig.~3 the time overhead of the SW-action is plotted.
The true time for a particular implementation 
should lie between the two limiting curves \cite{sw_hmc}. 
In any case, one notices that
for a situation where the number of iterations in the inversion
routine $N_{CG}$ exceeds roughly 250, the overhead is only about $20\%$.
One should remark that 
it is still an open question, how     
the SW-term can be implemented in a simulation program using 
the multiboson technique if one wants to keep 
heatbath or
over-relaxation methods for the updates. 

\section{Conclusion} 

Contrary to the situation 3 years ago, we now have two {\em exact}
methods to simulate dynamical fermions at our disposal which perform
competitively. The theoretical cost of the 
multiboson technique expressed in terms of $\lambda_{min}$ and $V$
is $V(logV)^2/\lambda_{min}^{3/2}$. This can be compared to the
cost of the molecular dynamics algorithms where one finds
$VV^{1/4}/\lambda_{min}^{7/4}$ for the HMC and 
$VV^{1/6}/\lambda_{min}^{3/2}$ for the Kramers equation algorithm.
Of course, one should keep in mind that the multiboson technique
has a large memory requirement and might therefore not be suitable
for particular machines. 

The real world situation of comparing the two kind of 
algorithms is summarized
in fig.~2. It demonstrates that for the points investigated
so far no clear priority for one or the other algorithm can be given.
It would certainly be important to have more data points. I also
consider it to be urgent to test the theoretical scaling behaviour 
as much as possible in order to see, whether the algorithms do what
we think they should do.

We should try to find out what is the magnitude of reversibility
violations in the HMC algorithm and attempt to quantify its implication
on physical observables. If lack of reversibility appears to be a problem
with the HMC method, one should use the multiboson technique or
the Kramers equation algorithm.

Dynamical fermion simulations with Symanzik improved actions have already
been 
started. Fig.~3 demonstrates that this can be done with
only a small overhead for simulations that need more than about
250 iterations for the matrix inversion. If also for dynamical
fermions improvement turns out to be important then this will
lead to substantial progress in the area of fermion simulation
algorithms.

\section*{Acknowledgement}
In the first place I would like to thank C. Liu for
a most enjoyable collaboration and 
many stimulating discussions. 
P. de Forcrand, A. Galli, I. Horvath, B. Jegerlehner, T. Kennedy,  
S. Meyer and T. Takaishi 
are gratefully acknowledged not only for very interesting
and helpful communications but also for making their results available to me
prior to publication. 
I am indebted to T. Lippert, M. L\"uscher, A. Slavnov and    
C. Thron
for many helpful discussions and comments.



\begin{thebibliography}{99}                                                     
%
\bibitem{tony}
A.~D.~Kennedy, Nucl.Phys.B (Proc. Suppl.) 30 (1993) 96.
%
\bibitem{martin}
M.~L\"uscher, Nucl. Phys. B418 (1994) 637.
%
\bibitem{bunk} 
B.~Bunk et.al., hep-lat/9411016; 
Nucl. Phys. B (Proc. Suppl.) 42 (1995) 49.
%
\bibitem{beat}
B.~Jegerlehner, hep-lat/9411065;
Nucl. Phys. B (Proc. Suppl.) 42 (1995) 879;
Nucl.Phys.B465 (1996) 487.
%
\bibitem{peardon}
M. Peardon, Nucl. Phys. B (Proc. Suppl.) 42 (1995) 891, hep-lat/9412008.
%
\bibitem{forcrand1}
A. Borici and P. de Forcrand, Nucl.Phys.B (Proc.Suppl.)47 (1996) 800,
hep-lat/9505021.  
%
\bibitem{forcrand2}
C. Alexandrou et.al.,
Nucl.Phys.B456 (1995) 296.                  
%
\bibitem{forcrand3}
P. de Forcrand, Nucl.Phys.B (Proc.Suppl.)47 (1996) 228,
hep-lat/9509082.
%
\bibitem{forcrand4}
A. Borrelli, P. de Forcrand and A. Galli, MPI Munich preprint,
hep-lat/9602016.
%
\bibitem{kramersboson}
K. Jansen, B.~Jegerlehner and C. Liu, Phys.Lett.B375 (1996) 255;
hep-lat/9604016.
%
\bibitem{borici}
A. Borici, ETH Z\"urich preprint, hep-lat/9602018.  
%
\bibitem{hmc}
S.~Duane et.al., 
Phys. Lett. B195 (1987) 216.
%
\bibitem{horowitz}
A.~M.~Horowitz, Phys. Lett. 156B (1985) 89; Nucl. Phys. B280 (1987) 510;
               Phys. Lett. 268B (1991) 247.
\bibitem{julius} J. Kuti, in ``Proceedings of the Thirty-Second Scottish
                 Universities Summer School in Physics'', St. Andrews (1987) 
                 311.
%
\bibitem{italiens}
M.~Beccaria, G.~Curci, Phys. Rev. D49 (1994) 2578;
M.~Beccaria, G.~Curci and L.~Galli, Phys. Rev. D49 (1994) 2590.
%
\bibitem{kramers} K. Jansen and C. Liu, Nucl. Phys. B453 (1995) 375.
%
\bibitem{andreas} A. Frommer, plenary talk given at this conference.  
%
\bibitem{revers}
R.~Edwards and A.~D.~Kennedy, private communications.
%
\bibitem{wuppertal1} U. Gl\"assner et. al., 
Nucl. Phys. B (Proc. Suppl.) 47 (1996) 386.
%
\bibitem{brower}
 R.~C. Brower et.al.,
Boston university preprint, BU-CCS-950602, hep-lat/9509012.
%
\bibitem{horvath} R. Edwards et.al., SCRI preprint, FSU-SCRI 96-49,
                  hep-lat/9606004.
%
\bibitem{liu} C. Liu, talk at this conference.
%
\bibitem{freekennedy} A.D. Kennedy et.al., hep-lat/9509043,
Nucl. Phys. B (Proc. Suppl.)47 (1996) 781.
%
\bibitem{meyer} S. Meyer, talk at this conference.
%
\bibitem{sf} M. L\"uscher et.al., 
Nucl.Phys.B 384 (1992) 168; 
S. Sint, Nucl.Phys.B421 (1994) 135, Nucl.Phys.B451 (1995) 416.
%
\bibitem{letter} K.~Jansen et.al.,
Phys.Lett.B372 (1996) 275.                     
%
\bibitem{paper} M.~L\"{u}scher et.al.,   
                DESY 96-086,
                hep-lat/9605038.     
%
\bibitem{wuppertal2} A. Frommer et.al., 
                     Int.J.Mod.Phys. C5 (1994) 1073.  
%
\bibitem{scales} J. Ortega, ``Introduction to Parallel and Vector 
                 Solution of Linear Systems'', Plenum, New York (1988).
%
\bibitem{kenpen}
A.~D.~Kennedy and B.~J.~Pendleton, Nucl. Phys. B (Proc.Suppl.) 20 (1991) 118. 
%
\bibitem{gupta} R. Gupta et.al., Phys.Rev.D40 (1989) 2072.
%
\bibitem{degrand} 
T.~Degrand and P.~Rossi, Comp. Phys. Comm. 60  (1990) 211.
%
\bibitem{sexy} 
J.~C.~Sexton and D.~H.~Weingarten, Nucl. Phys. B380 (1992) 665.
%
\bibitem{ssor} S. Fischer et.al., 
HLRZ preprint, HLRZ 4/96, hep-lat/9602019.
%
\bibitem{takaishi} T. Takaishi, talk at this conference.
%
\bibitem{gupta2} R. Gupta et.al., Phys.Rev.D44 (1991) 3272.
%
\bibitem{bermions} G.M. de Divitiis et.al., 
                   Nucl.Phys.B455 (1995) 274; 
                   Phys.Lett.B367 (1996) 279;
                   hep-lat/9603020;
                   hep-lat/9605002.  
%
\bibitem{shamir} V. Furman and Y. Shamir, Nucl.Phys.B439 (1995) 54. 
%
\bibitem{jaster} A. Jaster, DESY 96-083, hep-lat/9605011.
%
\bibitem{vranas} P. Vranas, talk at this conference.
%
\bibitem{montvay}
I. Montvay, Nucl.Phys.B466 (1996) 259.
%
\bibitem{hmd} A. Donini and M. Guagnelli, hep-lat/9605010
%
\bibitem{slavnov} A. Slavnov, preprint SMI-20-96, hep-lat/9605006.
%
\bibitem{thron} K.F. Liu and C. Thron, talks at this conference. 
%
\bibitem{taka} P. de Forcrand and T. Takaishi, hep-lat/9606020.
%
\bibitem{nf=1} P. de Forcrand, talk at this conference.
%
\bibitem{symanzik} K.~Symanzik, 
Nucl.Phys.B226(1987)187; 205.
%
\bibitem{clover} B. Sheikholeslami and R. Wohlert, Nucl. Phys. B259
 (1985) 572.
%
\bibitem{luo} X.-Q. Luo, Comp.Phys.Comm.94 (1996) 119.
%
\bibitem{sw_hmc} K. Jansen and C.Liu, DESY 96-042, 
                 hep-lat/9603008.  
%
\bibitem{wingate} M. Wingate, talk at this conference.
%
%
\bibitem{lippert} T. Lippert, talk at this conference.
%
\end{thebibliography}
\end{document}